\documentclass[11pt]{article}

\usepackage{amsxtra,amssymb,amsthm,amsmath,latexsym}
\usepackage{graphicx}
\usepackage{epsfig}
\usepackage{afterpage}

\newtheorem{lem}[]{Lemma}

 \newcommand{\lemref}[1]{Lemma~\ref{#1}}

\newcommand{\R}{{\mathbb R}}

\newcommand{\nb}{\nabla}

\newcommand{\dl}{{\delta}}
\newcommand{\bee}{\begin{equation*}}
\newcommand{\eee}{\end{equation*}}
\newcommand{\be}{\begin{equation}}
\newcommand{\ee}{\end{equation}}

\title{Electromagnetic wave scattering by many small particles
and creating materials with a desired permeability}
\author{A.G. Ramm \\
\small Department of Mathematics\\[-0.8ex]
\small Kansas State University, Manhattan, KS 66506-2602, USA\\
\small \texttt{ramm@math.ksu.edu}}

\date{}
\begin{document}

Progress in Electromagnetic Research, M, (PIER M) 14, (2010), 193-206.
\maketitle

\begin{abstract}
Scattering of electromagnetic (EM) waves by many small particles (bodies),
embedded in a homogeneous medium, is studied. Physical properties of
the particles are described by their boundary impedances. The
limiting equation is  obtained
for the effective EM field in the limiting medium, 
in the limit $a\to 0$, where $a$ is the characteristic size of a
particle and the number $M(a)$ of the particles tends to infinity at
a suitable rate. The proposed theory allows one to create a medium
with a desirable spatially inhomogeneous permeability. The main new 
physical
result is the explicit analytical formula for the permeability 
$\mu(x)$ of the limiting medium. While the initial medium
has a constant permeability $\mu_0$, the limiting medium, obtained 
as a result of embedding many small particles with prescribed boundary 
impedances, has a non-homogeneous permeability which is expressed 
analytically in terms of the density of the distribution of the
small particles and their boundary impedances. Therefore,
a new physical phenomenon is predicted theoretically, namely, appearance 
of a spatially inhomogeneous permeability as a result of embedding of many 
small particles whose physical properties are described by their boundary 
impedances.
\end{abstract}
 



\section{Introduction}
In this paper we outline a theory of electromagnetic (EM) wave
scattering by many small particles (bodies) embedded in a homogeneous 
medium which is described by the constant permittivity $\epsilon_0>0$,
permeability $\mu_0>0$ and, possibly, constant conductivity $\sigma_0\ge 
0$. The small particles are embedded in a finite domain $\Omega$. 
The medium, created by  the embedding of the small particles, has new 
physical properties. 
In particular, it has a spatially inhomogeneous magnetic permeability 
$\mu(x)$, which can be controlled by the choice of
the boundary impedances of the embedded small particles and their
distribution density.  This is 
a new physical effect, as far as the author knows.
An analytic formula for the permeability of the new medium is derived:
$$\mu(x)=\frac {\mu_0}{\Psi(x)},$$
where
$$\Psi(x)=1+\frac{8\pi}{3} i \epsilon_0 \omega h(x) N(x).$$ 
Here $\omega$ is the frequency of the EM field, $\epsilon_0$
is the constant dielectric parameter of the original medium, 
$h(x)$ is a function describing boundary impedances of the small
embedded particles, and $N(x)\geq 0$ is a function describing the 
distribution of these particles.  
We assume that in any subdomain $\Delta$, the number
$\mathcal{N}(\Delta)$ of the embedded particles $D_m$ is given by the
formula: 
$$\mathcal{N}(\Delta)=\frac{1}{a^{2-\kappa}}\int_{\Delta} 
N(x)dx[1+o(1)],\quad
a\to 0, $$ 
where $N(x)\geq 0$ is a continuous function, vanishing
outside of the finite domain $\Omega$ in which
small particles (bodies)  $D_m$ are distributed,
$\kappa\in(0,1)$ is a number one can choose at will, and
the boundary impedances of the small particles are defined by the
formula
$$\zeta_m=\frac{h(x_m)}{a^\kappa},\quad x_m\in D_m,$$ 
where $x_m$ is a point inside $m-$th particle $D_m$, Re $h(x)\geq 0$, and 
$h(x)$ is a continuous function vanishing 
outside $\Omega.$ The impedance boundary condition on the surface $S_m$
of the $m-$th particle $D_m$ is $E^t=\zeta_m [H^t,N]$, where $E^t$ ($H^t$)
is the tangential component of $E$ ($H$) on $S_m$, and $N$ is the unit
normal to $S_m$, pointing out of $D_m$.
 
Since one can choose the functions $N(x)$ and $h(x)$, one can create a 
desired magnetic permeability in $\Omega$. This is a novel idea, to the 
author's knowledge.

We also derive an analytic 
formula for the refraction coefficient of the medium in $\Omega$
created by the embedding of many small particles.
An equation for the EM field in the limiting medium is derived.
This medium is created  when the size $a$ of small particles tends to zero 
while the total number $M=M(a)$ of the particles tends to infinity at a 
suitable rate.

The refraction coefficient in the limiting medium is spatially 
inhomogeneous.

Our theory may be viewed as a "homogenization theory", but it differs
from the usual homogenization theory (see, e.g., \cite{CD}, \cite{MK},
and references therein) in several respects: we do not assume any periodic 
structure in the distribution of small bodies, our operators are 
non-selfadjoint, the spectrum of these operators is not discrete,
etc. Our ideas, methods, and techiques are quite different from the usual 
methods. 
These ideas are similar to  the ideas developed in
papers \cite{R509, R536}, where scalar wave scattering by small
bodies was studied, and in the papers \cite{R563},\cite{R598}.
However, the scattering of EM waves brought new technical difficulties
which are resolved in this paper. The difficulties come from the vectorial 
nature of the boundary conditions. Our arguments are valid for small
particles of arbitrary shapes. 

We also give a new numerical method for solving many-body wave-scattering
problems for small scatterers, see Section 5.2.

\section{EM wave scattering by many small particles}

We assume that many small bodies $D_m$, $1\leq
m\leq M$, are embedded in a homogeneous medium with constant
parameters $\epsilon_0$, $\mu_0$. Let $k^2=\omega^2\epsilon_0\mu_0$,
where $\omega$ is the frequency. Our arguments remain valid if one
assumes that the medium has a constant conductivity $\sigma_0>0$.
In this case $\epsilon_0$ is replaced by $\epsilon_0+i\frac 
{\sigma_0}{\omega}.$
Denote by $[E,H]=E\times H$ the cross 
product of two vectors, and by $(E,H)=E\cdot H$ the dot product of two 
vectors.

Electromagnetic (EM) wave scattering problem
consists of finding vectors $E$ and $H$ satisfying the Maxwell
equations: 
\be\label{e1} \nb \times E=i\omega\mu_0 H,\quad \nb\times
H=-i\omega \epsilon_0 E\quad \text{in } D:=\R^3\setminus
\cup_{m=1}^M D_m, \ee 
the impedance boundary conditions:
\be\label{e2} [N,[E,N]]=\zeta_m[H,N]\text{ on
} S_m,\ 1\leq m\leq M, \ee
and the radiation conditions:
\be\label{e3} E=E_0+v_E,\quad H=H_0+v_H,
 \ee
where $\zeta_m$ is the impedance, $N$ is the unit normal to $S_m$
pointing out of $D_m$, $E_0, H_0$ are the incident fields satisfying
equations \eqref{e1} in all of $\R^3$. One often assumes that
the incident wave is a plane wave, i.e., 
$E_0=\mathcal{E}e^{ik\alpha\cdot x}$, $\mathcal{E}$
is a constant vector, $\alpha\in S^2$ is a unit vector, $S^2$ is the
unit sphere in $\R^3$, $\alpha\cdot \mathcal{E}=0$, 
$v_E $ and $v_H$ satisfy the
radiation condition: $r(\frac{\partial v}{\partial
r}-ikv)=o(1)$ as $r:=|x|\to \infty$.

By impedance $\zeta_m$ we assume in this paper either a constant,
Re $\zeta_m\geq 0$, or a matrix function $2\times 2$ acting on the
tangential to $S_m$ vector fields, such that \be\label{e4}
\text{Re}(\zeta_mE^t,E^t)\geq 0\quad \forall E^t\in T_m, \ee where
$T_m$ is the set of all tangential to $S_m$ continuous vector fields
such that Div$E^t=0$, where Div is the surface divergence, and $E^t$
is the tangential component of $E$. 
Smallness
of $D_m$ means that $ka\ll 1$, where $a=0.5\max_{1\leq m\leq M}
\text{diam} D_m$. By the tangential to $S_m$ component $E^t$ of a vector
field $E$ the following is understood in this paper:
\be\label{e5}
E^t=E-N(E,N)=[N,[E,N]]. \ee This definition differs from the one
used often in the literature, namely, from the definition $E^t=[N,E]$.
Our definition \eqref{e5} corresponds to the geometrical meaning of
the tangential component of $E$ and, therefore, should be used. The
impedance boundary condition is written usually as
$$E^t=\zeta[H^t,N],$$ 
where the impedance $\zeta$ is a number. If one
uses definition \eqref{e5}, then this condition reduces to
\eqref{e2}, because $[[N,[H,N]],N]=[H,N].$

\begin{lem}\label{lem1}
Problem \eqref{e1}-\eqref{e4} has at most one solution.
\end{lem}
\lemref{lem1} is proved in Section 2.\\
Let us note that problem \eqref{e1}-\eqref{e4} is equivalent to the
problems \eqref{e6}, \eqref{e7}, \eqref{e3}, \eqref{e4}, where
\be\label{e6} \nb\times \nb\times E=k^2E\text{ in } D,\quad
H=\frac{\nb\times E}{i\omega \mu_0}, \ee \be\label{e7}
[N,[E,N]]=\frac{\zeta_m}{i\omega \mu_0}[\nb\times E,N]\text{ on }
S_m,\ 1\leq m\leq M. \ee 
Thus, we have reduced our problem to
finding one vector $E(x)$. If $E(x)$ is found, then
$H=\frac{\nb\times E}{i\omega\mu_0}.$ 

Let us look for $E$ of the
form \be\label{e8} E=E_0+\sum_{m=1}^M\nb \times
\int_{S_m}g(x,t)\sigma_m(t)dt,\quad
g(x,y)=\frac{e^{ik|x-y|}}{4\pi|x-y|}, \ee where $t\in S_m$ and $dt$
is an element of the area of $S_m$, $\sigma_m(t)\in T_m$. This $E$
for any continuous $\sigma_m(t)$ solves equation \eqref{e6} in $D$
because $E_0$ solves \eqref{e6} and 
\be\label{e9}\begin{split}
\nb\times\nb\times\nb\times \int_{S_m}g(x,t)\sigma_m(t)dt&=\nb \nb\cdot
\nb\times\int_{S_m}g(x,t)\sigma_m(t)dt\\
&-\nb^2\nb\times
\int_{S_m}g(x,t)\sigma_m(t)dt\\
&=k^2\nb\times \int_{S_m}g(x,t)\sigma_m(t)dt,\quad x\in D.
\end{split}\ee
Here we have used the known identity $div curl E=0,$ valid
for any smooth vector field $E$, and the known formula 
\be\label{eG}
-\nb^2 g(x,y)=k^2g(x,y)+\dl(x-y). 
\ee
The integral
$\int_{S_m}g(x,t)\sigma_m(t)dt$ satisfies the radiation condition.
Thus, formula \eqref{e8} solves problem \eqref{e6}, \eqref{e7},
\eqref{e3}, \eqref{e4}, if $\sigma_m(t)$ are chosen so that boundary
conditions \eqref{e7} are satisfied. 

Define the effective field
$E_e(x)=E_e^m(x)=E_e^{(m)}(x,a),$ acting on the $m-$th body $D_m$:
\be\label{e10} E_e(x):=E(x)-\nabla\times
\int_{S_m}g(x,t)\sigma_m(t)dt:=E_e^{(m)}(x), \ee where we assume
that $x$ is in a neigborhood of $S_m$, but $E_e(x)$ is defined for
all $x\in \R^3$. Let $x_m\in D_m$ be a point inside $D_m$, and
$d=d(a)$ be the distance between two neighboring small bodies. We assume
that
\be\label{e11} \lim_{a\to 0}\frac{a}{d(a)}=0,\quad \lim_{a\to
0}d(a)=0. \ee We will prove later that $E_e(x,a)$ tends to a limit
$E_e(x)$ as $a\to 0$, and $E_e(x)$ is a twice continuously
differentiable function. To derive an integral equation for
$\sigma_m=\sigma_m(t)$, substitute
$$E=E_e+\nb\times\int_{S_m}g(x,t)\sigma_m(t)dt$$ into \eqref{e7}, use
the formula 
\be\label{e12} [N,\nb\times
\int_{S_m}g(x,t)\sigma_m(t)dt]_{\mp}=\int_{S_m}
[N_s,[\nb_xg(x,t)|_{x=s},\sigma_m(t)]]dt\pm \frac{\sigma_m(t)}{2}, \ee 
(see, e.g., \cite{M}), the -(+) signs
denote the limiting values of the left-hand side of \eqref{e12} as $x\to 
s$ from
$D$ $(D_m)$, and get the following equation (see Appendix): 
\be\label{e13} \sigma_m(t)=A_m\sigma_m+f_m,\quad
1\leq m\leq M. \ee 
Here $A_m$ is a linear Fredholm-type integral
operator, and $f_m$ is a continuously differentiable function. Let
us specify $A_m$ and $f_m$. One has (see Appendix):
\be\label{e14}
f_m=2[f_e(s), N_s],\quad
f_e(s):=[N_s,[E_e(s),N_s]]-\frac{\zeta_m}{i\omega \mu_0}[\nb\times
E_e,N_s]. \ee 
Condition \eqref{e7} and formula \eqref{e12} yield
\be\label{e15}\begin{split}
&f_e(s)+\frac{1}{2}[\sigma_m(s),N_s]+
[\int_{S_m}[N_s,[\nb_sg(s,t),\sigma_m(t)]]dt,N_s]\\
&-\frac{\zeta_m}{i\omega
\mu_0}[\nb\times\nb\times\int_{S_m}g(x,t)\sigma_m(t)dt,N_s]|_{x\to
s}=0\end{split}\ee

Using the formula $\nb\times\nb\times =grad div  -\nb^2$, the relation
\be\label{e16}\begin{split}
\nb_x\nb_x \cdot \int_{S_m}g(x,t)\sigma_m(t)dt&=\nb_x\int_{S_m}(-\nb_t
g(x,t),\sigma_m(t))dt\\
&=\nb_x\int_{S_m}g(x,t)\text{Div}\sigma_m(t)dt=0, \end{split}\ee
where Div is the surface divergence, and the formula
\be\label{e17}
-\nb_x^2\int_{S_m}g(x,t)\sigma_m(t)dt=k^2\int_{S_m}g(x,t)\sigma_m(t)dt,\quad
x\in D, \ee 
where  equation  \eqref{eG}  was
used, one gets from \eqref{e15} the following equation
\be\label{e18} -[N_s,\sigma_m(s)]+2f_e(s)+2B\sigma_m=0. \ee 
Here
\be\label{e19} B\sigma_m:=[\int_{S_m}[N_s,[\nb_s
g(s,t),\sigma_m(t)]]dt,N_s]+\zeta_mi\omega
\epsilon_0[\int_{S_m}g(s,t)\sigma_m(t)dt,N_s]. \ee 
Take cross
product of $N_s$ with the left-hand side of \eqref{e18} and use the
formulas $N_s\cdot \sigma_m(s)=0$, $f_m:=f_m(s):=2[f_e(s), N_s]$, and 
\be\label{e20}
[N_s,[N_s,\sigma_m(s)]]=-\sigma_m(s), \ee 
to get from \eqref{e18}
equation \eqref{e13}: 
\be\label{e21}
\sigma_m(s)=2[f_e(s), N_s]-2[N_s,B\sigma_m]:=A_m\sigma_m+f_m, \ee
where $A_m\sigma_m=-2[N_s,B\sigma_m]$. The operator $A_m$ is linear and 
compact
in the space $C(S_m)$, so that equation \eqref{e21} is of Fredholm
type. Therefore, equation \eqref{e21} is solvable for any $f_m\in
T_m$ if the homogeneous version of \eqref{e21} has only the trivial
solution $\sigma_m=0$. In this case the solution $\sigma_m$ 
to equation \eqref{e21} is of the 
order of the right-hand side $f_m$, that is, $O(a^{-\kappa})$ as $a\to 
0$, see formula \eqref{e14}. Moreover, it follows from equation 
\eqref{e21} that the main term of the asymptotics of $\sigma_m$ 
as $a\to 0$ does not depend on $s\in S_m$.

\begin{lem}\label{lem2}
Assume that $\sigma_m\in T_m, $ $\sigma_m\in C(S_m)$, and
$\sigma_m(s)=A_m\sigma_m$. Then $\sigma_m=0$.
\end{lem}
\lemref{lem2} is proved in Section 2.

Let us assume that in any subdomain $\Delta$, the number
$\mathcal{N}(\Delta)$ of the embedded bodies $D_m$ is given by the
formula: 
\be\label{e22}
\mathcal{N}(\Delta)=\frac{1}{a^{2-\kappa}}\int_{\Delta}N(x)dx[1+o(1)],\quad
a\to 0, \ee 
where $N(x)\geq 0$ is a continuous function, vanishing
outside of a finite domain $\Omega$ in which 
small bodies  $D_m$ are distributed,
$\kappa\in(0,1)$ is a number one can choose at will. We also assume that 
\be\label{e23}
\zeta_m=\frac{h(x_m)}{a^\kappa},\quad x_m\in D_m, \ee where 
Re $h(x)\geq 0$, and $h(x)$ is a continuous function vanishing outside 
$\Omega.$ 

Let us write \eqref{e8} as \be\label{e24}
E(x)=E_0(x)+\sum_{m=1}^M[\nb_x g(x,x_m),Q_m]+\sum_{m=1}^M
\nb\times\int_{S_m}(g(x,t)-g(x,x_m))\sigma_m(t)dt, \ee where
\be\label{e25} Q_m:=\int_{S_m}\sigma_m(t)dt. 
\ee
Since $\sigma_m=O(a^{-\kappa})$, one has $Q_m=O(a^{2-\kappa})$. 
We want to prove
that the second sum in \eqref{e24} is negligible compared with the 
first sum. 
One has
\be\label{e26} j_1:=|[\nb_x g(x,x_m),Q_m]|\leq
O\left(\max\left\{\frac{1}{d^2},\frac{k}{d}\right\}\right)O(a^{2-\kappa}),
\ee 
\be\label{e27}
j_2:=|\nb\times\int_{S_m}(g(x,t)-g(x,x_m))\sigma_m(t)dt|\leq a
O\left(\max\left\{\frac{1}{d^3},\frac{k^2}{d}\right\}\right)O(a^{2-\kappa}),
\ee 
and 
\be\label{e28} \left| \frac{j_2}{j_1}\right|=O\left( \max
\left\{ \frac{a}{d},ka\right\}\right)\to 0,\qquad \frac a d=o(1),  \qquad 
a\to 0.  \ee
Thus, one may neglect the second sum in \eqref{e24}, and write
\be\label{e29} E(x)=E_0(x)+\sum_{m=1}^M[\nb_xg(x,x_m),Q_m] \ee 
with an error that tends to zero as $a\to 0$. 

Let us estimate $Q_m$
asymptotically, as $a\to 0$. Integrate equation \eqref{e21} over
$S_m$ to get 
\be\label{e30}
Q_m=2\int_{S_m}[f_e(s), N_s]ds-2\int_{S_m}[N_s,B\sigma_m]ds .\ee 
We will show in the Appendix  that the second term in the right-hand side 
of the above equation is equal to $-Q_m$ plus terms negligible compared 
with the first one as $a\to 0$. Thus, 
$$Q_m=\int_{S_m}[f_e(s), N_s]ds.$$
Let us estimate the first term.
It follows from equation \eqref{e14} that 
\be\label{e31}
[N_s,f_e]=[N_s,E_e]-\frac{\zeta_m}{i\omega \mu_0}[N_s,[\nb \times
E_e,N_s]]. \ee 
If $E_e$ tends to a finite limit as $a\to 0$, then
formula \eqref{e31} implies that \be\label{e32}
[N_s,f_e]=O(\zeta_m)=O\left(\frac{1}{a^\kappa}\right),\quad a\to 0.
\ee By \lemref{lem2} the operator $(I-A_m)^{-1}$ is bounded, so
$\sigma_m=O\left(\frac{1}{a^\kappa}\right)$, and \be\label{e33}
Q_m=O\left(a^{2-\kappa}\right),\quad a\to 0, \ee because integration
over $S_m$ adds factor $O(a^2)$. As $a\to 0$,
the sum \eqref{e29} converges to the integral \be\label{e34}
E=E_0+\nb\times\int_{\Omega}g(x,y)N(y)Q(y)dy, \ee where $Q(y)$ is
the function such that 
\be\label{e35} Q_m=Q(x_m)a^{2-\kappa}. \ee
The function $Q(y)$ can be expressed in terms of $E$: 
\be\label{e36}
Q(y)=-\frac{8\pi}{3} h(y)i\omega \epsilon_0 (\nb\times E)(y), \ee 
see Appendix.  
Here the factor $\frac{8\pi}{3}$ appears if $D_m$ are balls.
Otherwise a tensorial factor $c_{m}$, depending on the shape of $S_m$,
should be used in place of $\frac{8\pi}{3}$.

Thus, equation \eqref{e34} takes the form 
\be\label{e37} E(x)=E_0(x)-\frac{8\pi}{3}
i \omega \epsilon_0 \nb\times \int_{\Omega}g(x,y)\nb\times
E(y)h(y)N(y)dy. \ee
Let us derive physical conclusions from equation
\eqref{e37}. Taking $\nb\times\nb\times$ of \eqref{e37} 
yields
\be\label{e38}\begin{split} \nb\times\nb\times E&=k^2E_0(x)\\
&-\frac{8\pi}{3} i\omega \epsilon_0\nb\times (\text{grad
div}-\nb^2)\int_{\Omega}g(x,y)\nb\times E(y)h(y)N(y)dy\\
&=k^2E_0-k^2 \frac{8\pi}{3} i\omega \epsilon_0 
\nb\times\int_{\Omega}g(x,y) \nb\times E(y)h(y)N(y)dy\\
&-\frac{8\pi}{3}
i\omega\epsilon_0\nb \times(\nb\times E(x)h(x)N(x))\\
&=k^2E(x)-\frac{8\pi}{3} i \omega \epsilon_0 h(x)N(x) \nb\times\nb\times 
E\\
&-\frac{8\pi}{3} i \omega \epsilon_0[\nb (h(x)N(x)),\nb\times E(x)].
\end{split}\ee
Here we have used the known formula $\nb\times \text{grad}=0$, the
known equation  \eqref{eG}, and assumed for
simplicity that $h(x)$ is a scalar function. It follows from
\eqref{e38} that 
\be\label{e39} \nb\times\nb\times E=K^2(x)E-\frac{\frac{8\pi}{3} i \omega 
\epsilon_0}{1+\frac{8\pi}{3} i \omega \epsilon_0
h(x)N(x)}[\nb (h(x)N(x)),\nb\times E(x)], 
\ee 
where 
\be\label{e40}
K^2(x)=\frac{k^2}{1+\frac{8\pi}{3} i\omega\epsilon_0 h(x)N(x)},\quad
k^2=\omega^2 \epsilon_0 \mu_0. \ee 
If $\nb \times E=i\omega \mu (x) H$ and $\nb \times H=-i\omega 
\epsilon (x) E$, then 
\be\label{e41}\nb \times \nb \times E=\omega^2 \epsilon (x) \mu (x) E 
+[\frac {\nb \mu (x)}{\mu(x)}, \nb \times E].
\ee 
Comparing this equation with \eqref{e39}, one can identify the last
term in \eqref{e39} as coming from a variable permeability $\mu (x)$. 
This $\mu (x)$ appears in the limiting medium due to the boundary currents 
on the surfaces $S_m$, $1\le m \le M$. These currents appear because of 
the impedance boundary conditions \eqref{e7}.
Let us identify the permeability $\mu (x)$.  Denote 
$\Psi (x):=1+\frac{8\pi}{3} i \omega \epsilon_0 h(x) N(x)$. Let $\epsilon 
(x)=\epsilon_0$, $\epsilon_0= const,$ and define
$\mu(x):=\frac {\mu_0}{\Psi (x)}$. Then $K^2=\omega^2 
\epsilon_0 \mu (x)$,
and $\frac {\nb \mu (x)}{\mu(x)}=-\frac {\nb \Psi(x)}{\Psi (x)}$.
Consequently, formula \eqref{e39} has a clear physical meaning:
the electromagnetic properties of the limiting medium are described by the 
variable permeability:
\be\label{e42}
\mu(x)=\frac {\mu_0}{\Psi (x)}=\frac {\mu_0}{1+\frac{8\pi}{3} i \omega 
\epsilon_0 h(x) N(x)}.
\ee

\section{Conclusions}

{\it The limiting medium
is described by the new refraction coefficient $K^2(x)$ (see \eqref{e40}) 
and the new term
in the equation \eqref{e39}. This term is due to the spatially 
inhomogeneous permeability
$\mu(x)=\frac {\mu_0}{\Psi (x)}$ generated in the limiting medium
by the boundary impedances. 
The field $E(x)$ in the limiting medium
( and in equation \eqref{e39}) solves equation \eqref{e37}.

Therefore, we predict theoretically the new physical phenomenon:
by embedding many small particles with suitable boundary impedances 
into a given homogeneous medium, one can create a medium with a desired
spatially inhomogeneous permeability \eqref{e42}.

One can create material with a desired permeability $\mu(x)$
by embedding small particles with suitably chosen boundary impedances.
Indeed, by formula \eqref{e42} one can choose a complex-valued, in 
general, function $h(x)$, and a non-negative function  $N(x)\ge 0$,
describing the density distribution of the small particles,
so that the right-hand side of formula \eqref{e42} will yield a desired 
function $\mu(x)$.  }

\section{Proofs of Lemmas}
{\it Proof of \lemref{lem1}.}\\
\noindent From  equations \eqref{e1} one derives (the bar stands for 
complex conjugate): 
$$\int_{D_{R}}(\overline{H}\cdot \nb\times E-E\cdot \nb
\times \overline{H})dx=\int_{D_R}(i\omega \mu_0 |H|^2-i\omega
\epsilon_0|E|^2)dx,$$ where $D_R:=D\cap B_R$, and $R>0$ is so large that
$D_m\subset B_R:=\{x\ : \ |x|\leq R\}$ for all $m$. 
Recall that $\nb\cdot [E, \overline{H}]=\overline{H}\cdot\nb\times E
-E\cdot\nb\times \overline{H}$.
Applying the divergence theorem, using the radiation condition on the 
sphere $S_R=\partial B_R$, and taking real part, one gets
$$0=\sum_{m=1}^M\text{Re}\int_{S_m}[E,\overline{H}]\cdot N
ds=\sum_{m=1}^M\text{Re}\int_{S_m}\overline{\zeta_m}^{-1}\overline{E}^-_t
\cdot E^-_tds,$$
where $E^-_t$ is the limiting value of $E^t$ on $S_m$ from $D$,
$E^t=\zeta_m[H,N]$. This relation and assumption \eqref{e4} imply
$E^-_t=0$ on $S_m$ for all $m$. Thus, $E=H=0$ in $D$.\\
\lemref{lem1} is proved. \hfill $\Box$

{\it Proof of \lemref{lem2}.}

If $\sigma_m=A_m\sigma_m$, then the functions $H=\frac{\nb\times
E}{i\omega \mu_0}$ and $E(x)=\nb\times \int_{S_m}g(x,t)\sigma(t)dt$
solve equation \eqref{e1} in $D$, $E$ and $H$ satisfy the radiation
condition, and , condition \eqref{e2}. Thus, $E=H=0$ in $D$.
Consequently, 
\bee\begin{split} 0&=\nb\times\nb
\times\int_{S_m}g(x,t)\sigma_m(t)dt=(\text{grad
div}-\nb^2)\int_{S_m}g(x,t)\sigma_m(t)dt\\
&=k^2\int_{S_m}g(x,t)\sigma_m(t)dt,\quad x\in D.\end{split}\eee This
implies $\sigma_m(s)=0$.\\
\lemref{lem2} is proved. \hfill $\Box$

\section{Appendix}
In Section 5.1 equation \eqref{e37} is derived. In Section 5.2
a linear algebraic system (LAS) is derived for finding vectors $Q_m$
in equation \eqref{e35}.

5.1. Boundary condition \eqref{e7} yields \bee\begin{split}
0&=[N[E_e,N]]-\frac{\zeta_m}{i\omega \mu_0}[\nb\times
E_e,N]+[N,[\nb\times\int_{S_m}g(s,t)\sigma_m(t)dt,N]]\\
&-\frac{\zeta_m}{i\omega \mu_0}[\nb\times \nb\times
\int_{S_m}g(x,s)\sigma_m(t)dt,N].\end{split}\eee 
Let us denote
$$f_e:=[N,[E_e,N]]-\frac{\zeta_m}{i\omega\mu_0}[\nb\times E_e,N].$$
One has $\nb \times \nb \times=curl curl=grad div - \Delta$,
and 
\bee
\nb_x\cdot \int_{S_m}g(x,t)\sigma_m(t)dt=-\int_{S_m}\left(\nabla_t
g(x,t),\sigma_m(t) \right) dt=\int_{S_m} g(x,t)\nb_t\cdot
\sigma_m(t)dt=0, 
\eee 
and 
\bee
-\nb_x^2\int_{S_m}g(x,t)\sigma_m(t)dt=k^2\int_{S_m}g(x,t)\sigma_m(t)dt,
\eee because $-\nb_x^2 g(x,t)=k^2g(x,t)$, $x\neq t$, see \eqref{eG}. 
Thus, using
\eqref{e12}, one gets: \bee\begin{split}
0&=f_e+[\int_{S_m}[N_s,[\nb_s
g(s,t),\sigma_m(t)]]dt,N_s]+\frac{1}{2}[\sigma_m(s),N_s]\\
&+\frac{\zeta_m k^2}{i\omega
\mu_0}[N_s,\int_{S_m}g(s,t)\sigma_m(t)dt]. \end{split}\eee 
Cross multiply this by $N_s$ from the left and use the relation $N_s\cdot 
\sigma_m(s)=0$, to obtain
\bee\begin{split} 0&=[N_s, f_e]+[N_s,[\int_{S_m}[N_s,[\nb_s
g(s,t),\sigma_m(t)]]dt,N_s]]+\frac{1}{2}\sigma_m(s)\\
&-\zeta_m i\omega \epsilon_0[N_s,[N_s,\int_{S_m}g(s,t)\sigma_m(t)dt]].
\end{split}\eee
Note that 
\bee\begin{split} [N_s,[\int_{S_m}[N_s,[\nb_s
g(s,t),\sigma_m(t)]]dt,N_s]]&=\int_{S_m}[N_s,[\nb_s
g(s,t),\sigma_m(t)]]dt\\
&-N_s\int_{S_m}\left([N_s,[\nb_s
g(s,t), \sigma_m(t)]]dt,N_s\right)\\
&=\int_{S_m}[N_s,[\nb_s g(s,t),\sigma_m(t)]]dt.
\end{split},
\eee
where the integral before the last one vanishes because
its integrand vanishes as the dot product of two orthogonal vectors.
Consequently, 
\bee\begin{split} \sigma_m(t)&=2[f_e(s),N_s]+2\zeta_m
i\omega \epsilon_0 [N_s,[N_s,\int_{S_m}g(s,t)\sigma_m(t)dt]]\\
&-2\int_{S_m}[N_s,[\nb_sg(s,t),\sigma_m(t)]]dt:=A\sigma_m+f_m,
\end{split}\eee 
which is equation \eqref{e13}, and 
$$f_m:=2[f_e(s),N_s],$$
which is  equation \eqref{e14}. 

Denote
$$Q_m=\int_{S_m}\sigma_m(s)ds.$$ 
One has 
\bee\begin{split}
\int_{S_m}[[N_s,[E_e(s),N_s]],N_s]ds&=
\int_{S_m}[E_e(s),N_s]ds=
-\int_{D_m}\nb_x\times E_e dx,
\end{split}\eee
and 
\bee\begin{split} & \int_{S_m}[[\nb\times E_e,N_s],N_s]ds
=-\left( \int_{S_m}\nb\times E_e ds-\int_{S_m}N_s(\nb\times
E_e,N_s)ds\right)\\
&=-\int_{S_m}\nb\times E_e ds+\frac{4\pi a^2}{3}\nb\times E_e(x_m)\\
&=-\frac{8\pi a^2}{3}\nb\times E_e (x_m), \qquad a\to 0.
\end{split}\eee
Here we have used the formulas 
$$\int_{S_m}\nb\times E_e ds\sim 4\pi a^2 \nb\times E_e (x_m),\quad a\to 
0,$$
and
$$\int_{S}N_iN_jds=\frac{4\pi a^2}{3}\delta_{ij},$$
where $S$ is a sphere of radius $a$, $\{N_i\}_{i=1}^3$  are Cartesian
components of the outer unit normal to the sphere $S$, and 
$\delta_{ij}=0$ if $i\neq j$, $\delta_{ii}=1$.
 
Thus, 
\bee \int_{S_m}f_m(s)ds\sim -\frac{16\pi}{3}\zeta_m a^2
i\omega\epsilon_0 \nb\times E_e(x_m)=O(a^{2-\kappa}), \eee
as $a\to 0$,  provided that 
$$\zeta_m=\frac{h(x_m)}{a^\kappa},\qquad 0<\kappa< 1.$$
Let us now show that the term $\int_{S_m}A\sigma_m ds$ contributes 
$-Q$, so 
\be\label{eA.Q}
Q_m\sim 0.5\int_{S_m}f_m(s)ds, \qquad   a\to 0.
\ee
One has 
\bee\begin{split} &-2\int_{S_m}ds\int_{S_m}[N_s,[\nb_s
g(s,t),\sigma_m(t)]]dt\\
&=-2\int_{S_m}ds \int_{S_m}dt\left(\nb_s
g(s,t)(N_s,\sigma_m(t))-\sigma_m(t)\frac{\partial g(s,t)}{\partial
N_s} \right)
dt\\
&=-2\int_{S_m}ds \int_{S_m}dt \nb_s
g(s,t)(N_s,\sigma_m(t))+\int_{S_m}\sigma_m(t)dt 2\int_{S_m}ds
\frac{\partial g(s,t)}{\partial N_s}.
\end{split}\eee
Since 
$$2\int_{S_m}ds \frac{\partial g(s,t)}{\partial
N_s}=-2\int_{D_m}dxk^2g(x,t)-1,$$ 
one gets 
$$I:=\int_{S_m}dt
\sigma_m(t) 2\int_{S_m}ds\frac{\partial g(s,t)}{\partial
N_s}=-\int_{S_m}\sigma_m(t)dt-2k^2\int_{S_m}dt
\sigma_m(t)\int_{D_m}dx g(x,t).$$
Therefore
$$I:=-Q_m+I_1,$$
where the term $I_1$ is negligible compared with $I$.

If
$\int_{S_m}|\sigma_m(t)|dt<\infty$ and $\int_{S_m}\sigma_m(t)dt\neq 0$,
then 
$$|\int_{S_m}\sigma_m(t)dt|\gg |\int_{S_m}dt
\sigma_m(t)\int_{D_m}dx g(x,t)|,$$ 
because $|\int_{D_m}dx g(x,t)|=O(a^2)$ if $x\in D_m$. 

One has
$$ \big{|}-2\int_{S_m}ds\int_{S_m}dt 
\nb_sg(s,t)(N_s,\sigma_m(t))\big{|}\ll
\big{|}\int_{S_m}\sigma_m(t)dt\big{|}=|Q_m|,$$
because $|(N_s,\sigma_m(t))|=O(|s-t|)$ as $|s-t|\to 0$. 
Therefore,
\be\label{eA0}
Q_m=0.5\int_{S_m}\sigma_m(t)dt\sim -\frac{8\pi}{3}\zeta_m a^2
i\omega\epsilon_0 \nb\times E_e(x_m) ds,\quad a\to 0.
\ee
This yields the following formula (cf \eqref{e29}):
\be\label{eA1} E(x)=E_0(x)- \frac{8\pi}{3} i\omega\epsilon_0\sum_{m=1}^M
\zeta_m a^2[\nabla g(x,x_m), \nb\times E_e(x_m)], \qquad a\to 0.
\ee 
or, 
\be\label{eA2}
E(x)=E_0(x)-\frac{8\pi}{3} i\omega 
\epsilon_0\sum_{m=1}^Mh(x_m)a^{2-\kappa}
\left[ \nb_x g(x,x_m), \nb\times E_e (x_m)\right].
\ee 
Passing to the limit $a\to 0$, one obtains
\be\label{eA3}
E_e=E_0(x)-\frac{8\pi}{3} i \omega \epsilon_0\int_{\Omega} [\nb_x 
g(x,y),\nb \times E_e(y)] h(y)N(y)dy,
\ee
where $h(x)$ is the function in the formula $\zeta_m=\frac 
{h(x_m)}{a^{\kappa}}$, and $N(x)$ is the function in the definition
of $\mathcal{N}(\Delta)$.
The above passage to the limit is done by Theorem 1 from \cite{R598}, p. 
206. It uses the convergence of the collocation method for solving
equation \eqref{e37}, see \cite{R563}. 
Writing $E_e=E$ for the limiting field yields equation
\eqref{e37}.

5.2. In this Section a numerical method is developed for solving many-body
wave scattering problem when the scatterers are small in comparison with 
the wavelength. The method consists of a derivation of a linear algebraic 
system for finding vectors $\mathcal{P}_m:=(\nb\times E)(x_m)$,
$1\le m \le M.$ If 
$\mathcal{P}_m$ are found, then by formulas \eqref{e36} and \eqref{e35}
one finds 
\be\label{eA4}Q_m=-\frac{8\pi}{3}\pi i\omega \epsilon_0 h(x_m) 
a^{2-\kappa}\mathcal{P}_m,
\ee and, by formula \eqref{e29}, the field 
$E(x)$.
 
Let us derive linear algebraic system for finding
$\mathcal{P}_m$. 

Apply $\nb \times$ to equation \eqref{e29}, let $x=x_j$,
$1\le j \le M,$ and replace $\sum_{m=1}^M$ by the sum $\sum_{m\neq 
j, m=1}^M$. 

Then one obtains
\be\label{eA5}\mathcal{P}_j=\mathcal{P}_{0j}-\frac{8\pi}{3} i\omega 
\epsilon_0 a^{2-\kappa} \sum_{m\neq 
 j, m=1}^M (graddiv -\nabla^2)g(x,x_m)|_{x=x_j} h(x_m)\mathcal{P}_m,
\ee
where $1\le j \le M,$ and
\be\label{eA6}\mathcal{P}_{0j}:=(\nb \times E_0)(x_j).
\ee 
Equation \eqref{eA5} is a linear
algebraic system for finding $\mathcal{P}_m$. 
In the above derivation 
we have used the formula 
$$\nabla\times[A,B]=(B, \nabla)A-(A,\nabla)B+A (\nabla,B)- B(\nabla,A),$$
which yields
$$\nabla\times [\nabla g,Q_m]=(Q_m,\nabla)\nabla g-Q_m\nabla^2 g,$$
where
$$(Q_m,\nabla)\nabla g=\sum_{i=1}^3 e_i\sum_{i'=1}^3 
Q_{mi'}\frac{\partial^2 g}{\partial x_{i}\partial x_{i'}},\quad 
Q_{mi}=(Q_m,e_i),$$ 
and $e_i$, $i=1,2,3,$ is the orthonormal Cartesian basis of $\R^3$.


\newpage
\begin{thebibliography}{00}

\bibitem{CD} D. Cioranescu, P. Donato, {\it An introduction to 
homogenization}, Oxford Univ. Press, New York, 1999.

\bibitem{MK} V.Marchenko, E. Khruslov, {\it  Homogenization of partial 
differential equations,} Birkh\"auser, Boston, 2006.

\bibitem{M} C. M\"{u}ller, {\it Foundations of the 
mathematical theory of
electromagnetic waves}, Springer-Verlag, Berlin, 1969.


\bibitem{R509}  A.~G.~Ramm, Many-body wave scattering by small bodies 
and
applications, J. Math. Phys., 48, N10, (2007), 103511.

\bibitem{R536} A.~G.~Ramm, Wave scattering by many small particles
embedded in a medium, Phys. Lett. A, 372/17, (2008), 3064-3070.

\bibitem{R563} A.~G.~Ramm, A collocation method for solving integral 
equations, Internat. Journ. Comp. Sci. Math (IJCSM), 3, N2, (2009),
222-228. 

\bibitem{R598}
A.~G.~Ramm,  Electromagnetic wave
scattering by many small bodies and creating materials
with a desired refraction coefficient,
Progress in Electromagnetic Research M (PIER M), 13, (2010), 203-215.



\end{thebibliography}
\end{document}